\documentclass[twocolumn,amsmath,amssymb,prl]{revtex4}

\usepackage{graphicx}
\usepackage{dcolumn}
\usepackage{bm}

\begin{document}



\title{A high pressure calorimetric experiment to validate the liquid-liquid critical point
hypothesis in water}

\author{Manuel I. Marqu\'es}
\email{manuel.marques@uam.es}

\affiliation{%
~Departamento de F\'isica de Materiales C-IV,
Universidad Aut\'onoma de Madrid, 28049 Madrid, Spain\\
}%



\begin{abstract}
An experimental proposal to test the existence of a liquid-liquid critical point 
in water, based on high pressure calorimetric measurements ,is presented 
on this paper. Considering the existence of an intramolecular correlation
in the water molecule 
we show how the response of the specific heat at high pressure is 
different depending on the existence, or not, of the second critical point.
If the liquid-liquid critical point hypothesis is true there must be a
maximum in the specific heat at some temperature $T>T_{H}$
for any pressure $P>P_{c}$ (being $T_{H}$ the homogeneous nucleation temperature
and $P_{c}$ the pressure of the second critical point). This maximum does not appear
for the singularity free scenario. 
\end{abstract}


\maketitle


Liquid water exists in a metastable supercooled state far bellow the melting temperature.
A number of thermodynamic response functions of supercooled water such as the isothermal
compressibility and the constant pressure specific heat, show a power-law divergence at T=228K
and P=1atm \cite{Speedy,Angell}. In order to explain these anomalies several 
theories have been proposed. 

The liquid-liquid phase transition hypothesis \cite{Poole} proposes the 
existence of a first order line of phase transitions separating two liquid states of different densities:
the high density liquid (HDL) and the low density liquid (LDL) state. This line of phase transitions has a negative
slope in the $P-T$ phase diagram and ends up in a critical point at $T_{c}\sim200$K and $P_{c}\sim1.7$kbars.
Another possible scenario is singularity-free \cite{Stanley}. In this proposal the anomalies
found in the experiments are not considered to end up in real singularities but in non divergent maxima due 
to the anticorrelated fluctuations of volume and entropy. 

The direct observation of a possible first order liquid-liquid phase transition line 
and a second critical point in water has been hampered by the intervention of the homogeneous nucleation process
which takes place at higher temperatures $T_{H}(P)$  than the ones corresponding to the hypothesized coexistence line. 
In this paper we are going to focus our attention in the following question: Can we expect any difference (at temperatures $T>T_{H}(P)$) 
in the calorimetric response of supercooled water depending on the existence, or not, of a second critical point? 

In order to give a possible answer to this question we are going to consider a liquid model where the existence, or not, of a second liquid-liquid critical
point is easily tunable. The model is described in detail in \cite{Franzese}. In the liquid phase, each water molecule interacts with another four water
molecules forming hydrogen bonds with energy $-J$. The formation of a hydrogen bond increases the volume of the system by a certain amount
 $\delta V$. In the simplest version of this model,
there is no correlation between the possible four hydrogen bonds formed by a particular water molecule. The correlation length related to the
formation of the hydrogen bond network does not diverge 
and the thermodynamic response functions present a simple (non-divergent) maximum at a temperature $T_{h}(P)$.  
This temperature, $T_{h}(P)$, is always proportional to the absolute value of the interaction energy due to the formation of a hydrogen bond 
at a given pressure, $T_{h}(P) \propto (J-(P\delta V)$) \cite{Franzese}. Note that $T_{h}(P) \propto J$ for $P=0$ and $T_{h}(P)=0$ for $P=J/\delta V$.
The behavior of this simple model is the one described by the singularity free scenario. 

However
by considering an internal correlation among the four hydrogen bonds formed by each single molecule the scenario changes from singularity-
free to critical point like \cite{Franzese}. This internal degree of correlation in the water molecules is tuned by a new intramolecular term 
with energy $-J_{\sigma}$ added
to the Hamiltonian of the liquid \cite{Franzese}. If $J_{\sigma}\rightarrow\infty$ the water molecule is always internally correlated and, for any value of the
pressure,  the thermodynamic functions are always divergent at $T_{h}(P)$.

The liquid-liquid critical point at $T_{c}$ and $P_{c}$ shows up when $J_{\sigma}$ is finite and smaller than $J$. In this case, the water molecules
are able to correlate themselves internally only if $T<T_{*}$, being $T_{*}$ a pressure independent temperature proportional to $J_{\sigma}$.  
For $P<P_{c}$ the water molecules
are not internally correlated at $T_{h}(P)$  (i.e. $T_{h}(P)>T_{*}$), the correlation length of the hydrogen bonds network is unable to tend to infinity, 
and the response of the 
thermodynamic functions is non-divergent. 

However, as the pressure increases $T_{h}(P)$ decreases. So, there is a critical value of the pressure $P=P_{c}$ where the water
molecules are internally correlated at $T_{h}(P_{c})$ (i.e. $T_{h}(P_{c}) \leq T_{*}$), the correlation length of the hydrogens bond network is able to go to infinity, 
and the response of
the thermodynamic functions is divergent at $T_{h}(P_{c})=T_{c}$. Of course, for any $P>P_{c}$, the four hydrogen bonds
formed by any water molecule are always internally ordered at  $T_{h}(P)<T_{*}$ and the response of the thermodynamic functions are going to be 
the ones corresponding
to a first order phase transition between two liquids with different density.

This mechanism is explained in detail on Fig.1.
Note how, by tunning the intra-molecular term $J_{\sigma}$, we can easily change the physical mechanism of the simulated liquid
 form singularity free ($J_{\sigma}=0$) to 
liquid-liquid phase transition ($0 \neq J_{\sigma}<J$).

In real water, this intra-molecular term in the water molecule could be related to the value of the H-O-H angle. Experiments show that the relative 
orientations of the arms in the water molecule are correlated, with the average
H-O-H angle equal to $104.45^{o}$ in the isolated molecule, $104.474^{o}$ in the gas and $106^{o}$ in the high $T$ liquid \cite {Kern}, suggesting an 
intramolecular interaction between the arms. This interaction must be finite because the angle changes with $T$, 
consistent with ab-initio calculations \cite {Silvestrelli} and molecular dynamics simulations \cite {Netz}. 

\begin{figure}
\includegraphics[width=7cm,height=10cm,angle=0]{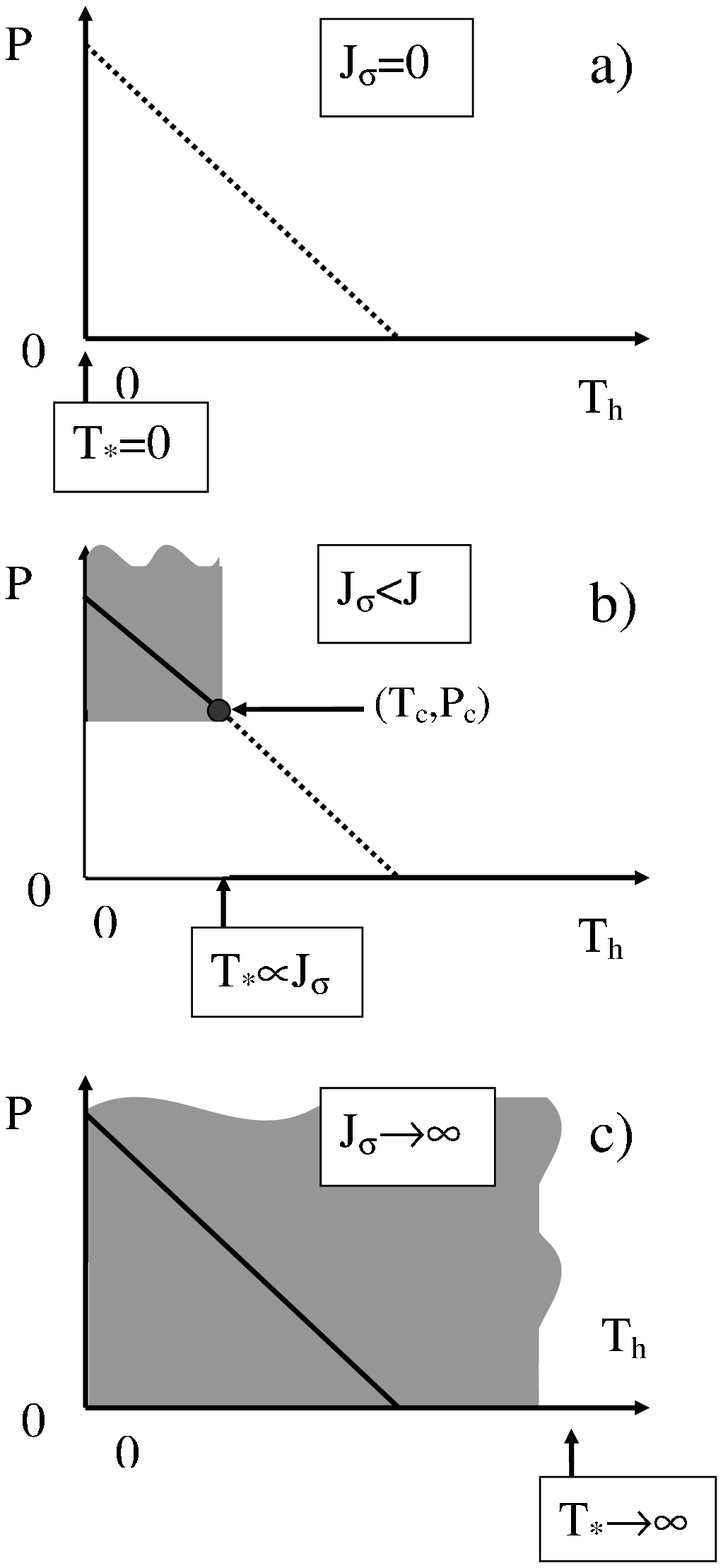}
\caption{Three possible scenarios depending on the value of $J_{\sigma}$. (a) If $J_{\sigma}=0$ 
there is no internal correlation in the water molecule for any pressure 
($T_{*}=0$) and there is no long range correlation in the formation of the hydrogen bonded network. The transition is always singularity-free (dotted line). 
(b) If $J_{\sigma}<J$ and different from zero, there is long range correlation in the transition to a low density liquid only when the transition 
temperature $T_{h}$ is smaller than the temperature $T_{*}$ corresponding to the internal correlation of the water molecules. In this particular case  
we obtain a critical point and a first order phase transition for any pressure $P>P_{c}$ and for any temperature $T<T_{c} \sim T_{*}$ 
(gray region and continuous line). (c) If $J_{\sigma}$ is large enough compared to $J$  there is always internal correlation in
the water molecules for any pressure and the transition to a low density liquid is first order for any pressure (gray region and continuous line).    
}
\label{fig1}
\end{figure}

Based on the model, we propose the following hypothesis: If there is a second critical point at $P_{c}$ and we cool down our liquid deep into the
supercooled region at constant $P>P_{c}$, there is going to be internal correlation inside the water molecule at a temperature $T_{*}>T_{h}$ 
(i.e. before homogeneous nucleation takes place).     
The fluctuations on the energy of the system due to this intra-molecular ordering give rise to a maximum in the specific heat with the following
characteristics:(i) It is located at a temperature larger than the one corresponding to nucleation ($T_{*}>T_{h}$), 
(ii) it is non-divergent (because it is not due to the cooperative effect of many molecules but to the internal ordering of each molecule 
separately)
and (iii) the position of the maximum is going to be pressure independent (i.e. since $J_{\sigma}$ is constant and does not depend on pressure 
the value of $T_{*}$ should be also pressure independent). 

On the other hand if $P<P_{c}$ and we cool down our liquid, nucleation takes place before intra-molecular ordering
($T_{*}<T_{h}$) and no maximum appears in the specific heat (only the regular nucleation maximum reachable in numerical simulations but 
unreachable in a real supercooled bulk water experiments).        
In the case of a singularity free scenario where there is no critical point (critical pressure), there is never any internal ordering in the water molecules and,
independently of the pressure applied, no maximum in the specif heat is found.

In order to test this hypothesis we have simulated the liquid phase of the model described in Ref. \cite{Franzese} by computing the total energy density
of states \cite{Wang}. We have chosen the following values for the parameters: $J/\epsilon=50$ (being $\epsilon$ the
value of the van der Waals interaction energy which in our case, since we are only considering the condensed liquid phase, does not play any role) 
and $\delta V/v_{o}=50$
(being $v_{o}$ the hard core volume of the water molecule). We have performed two sets of calculations, one for a liquid with no liquid-liquid critical point 
($J_{\sigma}/\epsilon=0$) and another one with a liquid-liquid critical point ($J_{\sigma}/\epsilon=5$). All other parameters are the same than in 
Ref. \cite{Franzese}. The behavior of energy fluctuations (constant pressure specific heat) has been calculated for a system of $225$ water molecules with
 periodic boundary conditions and for the following values of the pressure $Pv_{o}/\epsilon=0.54,0.74,0.84,0.89,0.94,0.955,0.97$. Results are shown in Fig.2

\begin{figure}
\includegraphics[width=7cm,height=7cm,angle=-90]{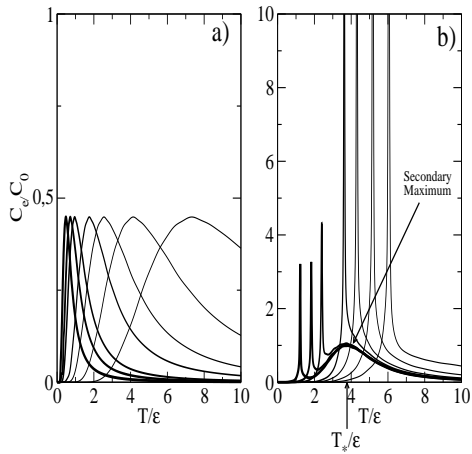}
\caption{Constant pressure specific heat vs. temperature for different pressures $Pv_{o}/\epsilon=$0.54 (thiner line),0.74,0.84,0.89,0.94,0.955,0.97 (thicker line) and for
(a) $J_{\sigma} / \epsilon =0$ (singularity free scenario) and (b) $J_{\sigma} / \epsilon=5$ (liquid-liquid critical point scenario). Note how we find secondary maxima only in (b) and
for  $Pv_{o}/\epsilon>0.89$, indicating that the critical pressure of the model must be close to $Pv_{o}/\epsilon=0.89$. The relative values of the maxima of the
peaks are system-size dependent expect for the secondary peaks indicated in (b). Absolute specific heat values are normalized to $C_{0}$, the value of the
secondary maxima found in (b). 
}
\label{fig2}
\end{figure}

Fig.2 (a) are the results for the liquid with no-critical point. In
this case no secondary maximum is found at any pressure. 
Fig.2 (b) are the results for the liquid with a critical point and a coexistence line between a LDL phase and a HDL phase. 
Note how the specific heat, for pressures $Pv_{o}/\epsilon>0.89$, has
two maxima. The characteristics of the secondary peaks are the following: they are located always at a temperature ($T_{*}$) larger than the nucleation
temperature, they are non-divergent, and the temperature  ($T_{*}$) is pressure independent.  These findings agree with our hypothesis previously presented 
and suggest that there is a critical point at a pressure $Pv_{o}/\epsilon\sim0.89$ for liquid (b). 

There is an important comment about the values of the main maxima of the specific heat 
shown in Fig.2. Since
we are dealing with an small system with finite size effects the concrete value of the maxima are not representative. For example, for the isothermal 
compressibility, it is found that maxima are proportional to the number of molecules in the first order phase transition, and that they scale as a power of the
number of molecules when the transition is at the critical point (second order phase transition) \cite {Franzese}. In any case, they are
not measurable in a real experiment due to the homogeneous crystallization of water. On the contrary, the secondary maxima shown on Fig.2 (b) are 
clearly non-divergent because they are size independent (not shown). 

\begin{figure}
\includegraphics[width=9cm,height=9cm,angle=-90]{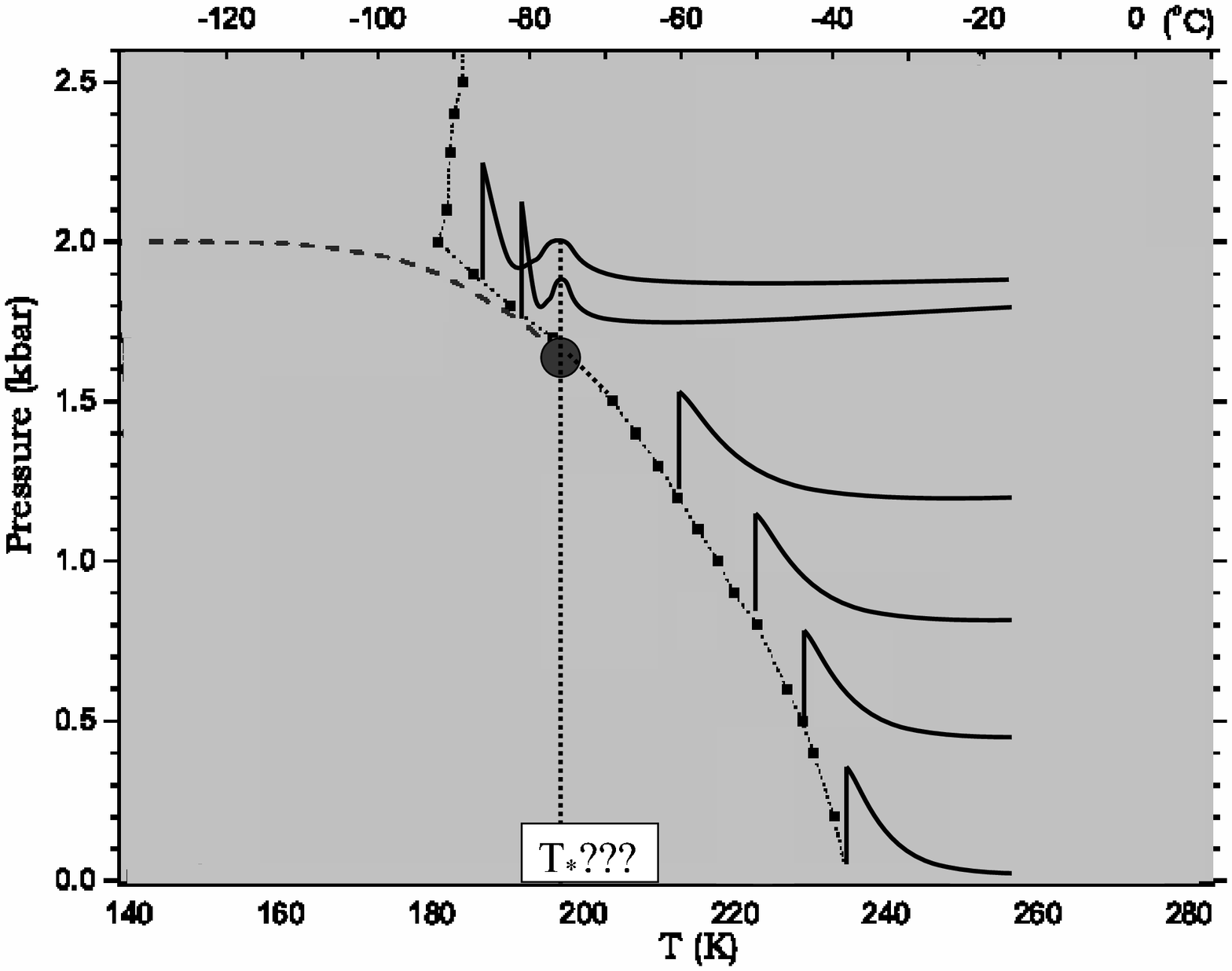}
\caption{Schematic representation of the experiment proposed to test the existence of a second liquid-liquid critical point
at $P_{c} \sim1.7$ kbar and $T_{c} \sim200K$ (black circle) . Constant pressure specific heat measurements should
be performed by cooling down water until reaching the homogeneous nucleation temperature (squares). For pressures
smaller than $\sim 1.7$ kbar the specific heat should present the usual constant power-law increase when cooling in the deeply 
supercooled region.
However, for larger pressures (in the range form 1.7kbar to 2kbar), there should be a previous non-divergent maximum in 
$T=T_{*}$ and close to the temperature of 200K independently of the pressure applied. This behavior for the specific heat has been 
schematically represented on the figure (thick continuous lines). Dotted line is the proposed coexistence line. 
}
\label{fig3}
\end{figure}

This non-divergent secondary peak found for $P>P_{c}$ could be measured in a real, high pressure, water experiment, since it is located
at temperatures above the nucleation temperature. The scheme of the concrete experiment to be performed is presented on Fig.3. Calorimetric measurements
on water should be performed for pressures ranging from atmospheric pressure to approximately $P=2$ kbar in the supercooled region. Until now,
calorimetric measurements in bulk water have been mainly performed at low pressures \cite{Debenedetti} and no maximum has been found down
to $T=240$K. However, following our hypothesis, if there is a liquid-liquid critical point, calorimetric measurements should present a non-divergent 
maxima for values 
of the pressure $P>P_{c}\sim 1.7$kbar \cite{Liu} at a temperature $T_{*}$ above the homogeneous nucleation temperature (which is
located between 200K and 180K depending on the value of the pressure applied). The value of $T_{*}$ should be almost pressure independent
and close to the hypothesized critical temperature $T_{c}\sim 200K$ \cite {Liu}.
This secondary maximum should not show up in other quantities such as the compressibility because the intra-molecular term we are considering for the
water molecule does not imply any change in volume. On the other hand, if the real scenario in water is singularity-free there should be no maxima 
in the specific heat. 

To conclude, the liquid-liquid critical point and the existence of an internal correlation in the water molecule could be strongly supported by the 
existence of non-divergent maxima in the water specific heat at high pressures. Of course, an important question is: in case these experiments are 
performed and no maxima are found for any pressure, does it immediately implies that there is no second critical point in water? The answer is no, because
there is also the possibility of the existence of a liquid-liquid critical point in water that is not related to an internal correlation inside the 
water molecule due to the existence of an intra-molecular term in the Hamiltonian.

\end{document}